\documentclass[pdflatex,sn-mathphys-num]{sn-jnl}
\usepackage{amsmath,amssymb,amsfonts}
\usepackage{graphicx}
\usepackage{tikz}
\usepackage{booktabs}
\usepackage{placeins}
\usepackage{etoolbox}
\AtBeginEnvironment{thebibliography}{\interlinepenalty=10000}
\usepackage[T1]{fontenc}
\usepackage{microtype}
\usepackage[title]{appendix}
\hypersetup{hidelinks,bookmarksnumbered=true,bookmarksopen=true,
  pdftitle={Hexagonal Stacking Maximizes Proton Configurational Entropy among Ice-I Polytypes},
  pdfauthor={Zhengyue Chen and Sheng Ran},
  pdfkeywords={Residual entropy, Ice-I polytypes, Transfer matrices, Rigorous bounds, Eulerian orientations, Configurational entropy},
  pdfsubject={Manuscript for Journal of Statistical Physics}}
\newcommand{\Ic}{\mathrm{Ic}}
\newcommand{\Ih}{\mathrm{Ih}}
\begin{document}
\title[Proton configurational entropy of ice-I polytypes]{Hexagonal Stacking Maximizes Proton Configurational Entropy among Ice-I Polytypes}
\author[1,2]{\fnm{Zhengyue} \sur{Chen}}\email{czhengyue@wustl.edu}
\author*[1]{\fnm{Sheng} \sur{Ran}}\email{rans@wustl.edu}
\affil[1]{\orgdiv{Department of Physics}, \orgname{Washington University in St. Louis}, \orgaddress{\city{St. Louis}, \state{MO}, \postcode{63130}, \country{USA}}}
\affil[2]{\orgdiv{Institute of Materials Science \& Engineering}, \orgname{Washington University in St. Louis}, \orgaddress{\city{St. Louis}, \state{MO}, \postcode{63130}, \country{USA}}}

\abstract{Ice I admits cubic, hexagonal, and mixed layer stackings, but rigorous comparisons of its
proton configurational entropy have focused on the two ideal endmembers. We identify every
cyclic uniform-registry ice-I stacking with a word in a nonnegative layer-transfer operator
and its transpose. Applying the Schatten--H\"older inequality to every even-length word proves
that alternating hexagonal stacking maximizes the ice-rule configuration count at each common
finite cross-section. Consequently, hexagonal ice maximizes the thermodynamic configuration
constant among all periodic uniform-registry ice-I polytypes. We also construct rigorous
lower and upper bounds without finite-size extrapolation. Restricting Nagle's positive
even-subgraph expansion to disjoint blocks gives the lower bound; Finner's hypergraph
H\"older inequality and rational Collatz--Wielandt certificates for two-replica prism operators
give the upper bounds. The resulting configuration constant $w$, whose logarithm is the
entropy per molecule in units of Boltzmann's constant, satisfies
$1.503360\le w\le1.540196$, with the sharper cubic-ice ceiling $w\le1.527699$.
The same bracket controls lower and upper configuration growth rates of fixed aperiodic stackings along
the specified prism-tileable exhaustions. These results establish hexagonal stacking as a
possibly nonunique maximizer; equality of the cubic and hexagonal entropy constants remains
unresolved.}


\maketitle

\section{Introduction}\label{sec:introduction}

Even as the temperature approaches zero, proton-disordered ice retains a measurable residual
entropy \cite{GiauqueStout1936,Pauling1935}. Its microscopic origin is configurational: the
tetrahedral oxygen network can be treated as fixed while each proton occupies one of two
positions along the bond between neighboring oxygens. These local choices obey the
Bernal--Fowler ice rule---every oxygen has two nearby protons and two more distant protons---but
the rule still permits exponentially many global arrangements. Hydrogen is the lightest
element, and once the much heavier oxygen framework is fixed, proton placement becomes the
natural local degree of freedom. Ice therefore reduces a molecular solid to a simple discrete
model of how local constraints generate macroscopic degeneracy. The quantity counted below is
the proton configurational entropy generated by these ice-rule arrangements.

We focus on ice I because its polytypes provide a controlled way to isolate the effect of crystal
stacking. Hexagonal ice $\Ih$, cubic ice $\Ic$, and stacking-disordered ice are assembled from
the same tetrahedrally coordinated layers and obey the same local ice rule; they differ in how
those layers are registered. Hexagonal ice follows $ABAB\ldots$, cubic ice follows
$ABCABC\ldots$, and mixed polytypes use nonuniform sequences of the same two registry steps
(Fig.~\ref{fig:stacking}).
Ice crystallized from supercooled water is commonly stacking-disordered
\cite{Malkin2012,Malkin2015}; nearly defect-free cubic ice has only recently been isolated
\cite{delRosso2020,Komatsu2020}; and stacking faults have been imaged directly
\cite{Huang2023}. Other ice phases need not preserve this common oxygen network, so comparing
them would change the local structure together with the stacking geometry. Restricting to ice I
leaves a simpler question: when the local rule and layer building blocks are held fixed, can the
global stacking sequence change the number of allowed proton configurations?

\begin{figure}[!tbp]
\centering
\resizebox{\textwidth}{!}{%
\begin{tikzpicture}[scale=1.0,transform shape,
  site/.style={circle,fill=black,inner sep=1.15pt},
  lay/.style={line width=0.9pt},
  bond/.style={line width=0.6pt,gray!70},
  lbl/.style={font=\small}]
 \def\layer#1#2#3{%
   \foreach \k in {0,...,5}{
     \pgfmathsetmacro{\xx}{#1+\k*0.62}
     \pgfmathsetmacro{\zz}{#2+ifthenelse(mod(\k,2)==0,0,0.19)}
     \node[site] (#3\k) at (\xx,\zz) {};}
   \foreach \k in {0,...,4}{\draw[lay] (#3\k)--(#3\the\numexpr\k+1\relax);}}
 \begin{scope}
   \layer{0}{0}{a}   \layer{0.21}{1.15}{b}  \layer{0.42}{2.30}{c}  \layer{0}{3.45}{d}
   \foreach \p/\q in {a/b,b/c,c/d}{\foreach \k in {0,2,4}{\draw[bond] (\p\k)--(\q\k);}}
   \node[lbl,anchor=east] at (-0.25,0.1)  {A};
   \node[lbl,anchor=east] at (-0.25,1.25) {B};
   \node[lbl,anchor=east] at (-0.25,2.40) {C};
   \node[lbl,anchor=east] at (-0.25,3.55) {A};
   \foreach \y in {0.62,1.77,2.92}{\draw[->,line width=0.8pt] (3.55,\y-0.05)--(3.55,\y+0.42);}
   \foreach \y in {0.85,2.00,3.15}{\node[lbl,anchor=west] at (3.62,\y) {$K$};}
   \node[lbl] at (1.6,-0.75) {cubic ice \textsf{Ic}: ABC, one handedness};
   \node[lbl] at (1.6,-1.15) {word $K^{\,n}$};
 \end{scope}
 \begin{scope}[xshift=6.6cm]
   \layer{0}{0}{p}   \layer{0.21}{1.15}{q}  \layer{0}{2.30}{r}  \layer{0.21}{3.45}{s}
   \foreach \pp/\qq in {p/q,q/r,r/s}{\foreach \k in {0,2,4}{\draw[bond] (\pp\k)--(\qq\k);}}
   \node[lbl,anchor=east] at (-0.25,0.1)  {A};
   \node[lbl,anchor=east] at (-0.25,1.25) {B};
   \node[lbl,anchor=east] at (-0.25,2.40) {A};
   \node[lbl,anchor=east] at (-0.25,3.55) {B};
   \foreach \y in {0.62,1.77,2.92}{\draw[->,line width=0.8pt] (3.55,\y-0.05)--(3.55,\y+0.42);}
   \node[lbl,anchor=west] at (3.62,0.85) {$K$};
   \node[lbl,anchor=west] at (3.62,2.00) {$K^{\mathsf T}$};
   \node[lbl,anchor=west] at (3.62,3.15) {$K$};
   \node[lbl] at (1.6,-0.75) {hexagonal ice \textsf{Ih}: AB, alternating};
   \node[lbl] at (1.6,-1.15) {word $(KK^{\mathsf T})^{m}$};
 \end{scope}
 \begin{scope}[xshift=3.05cm,yshift=-2.60cm]
   \node[lbl,anchor=east] at (-0.35,0) {any mixed stacking};
   \foreach \i/\r in {0/A,1/B,2/C,3/B,4/C,5/B,6/A}{
     \draw[lay] (\i*1.5,-0.16)--(\i*1.5+0.62,-0.16);
     \node[lbl,anchor=south] at (\i*1.5+0.31,-0.10) {\r};}
   \foreach \i/\op in {0/{K},1/{K},2/{K^{\mathsf T}},3/{K},4/{K^{\mathsf T}},5/{K^{\mathsf T}}}{
     \draw[->,line width=0.7pt] (\i*1.5+0.72,-0.16)--(\i*1.5+1.4,-0.16);
     \node[lbl,anchor=south,inner sep=1.5pt] at (\i*1.5+1.06,-0.13) {$\op$};}
   \node[lbl,anchor=north] at (4.65,-1.05)
     {$Z_Q(\sigma)=\operatorname{tr}\big(KKK^{\mathsf T}KK^{\mathsf T}K^{\mathsf T}\big)
       \ \le\ Z_Q(\Ih)$};
 \end{scope}
 \begin{scope}[xshift=12.3cm,yshift=1.55cm]
   \node[circle,fill=black,inner sep=1.6pt] (O) at (0,0) {};
   \foreach \a/\d in {60/in,180/in,300/out,20/out}{}
   \draw[->,line width=0.85pt] (135:1.0)--(135:0.24);
   \draw[->,line width=0.85pt] (215:1.0)--(215:0.24);
   \draw[->,line width=0.85pt] (0.24,0.06)--(35:1.0);
   \draw[->,line width=0.85pt] (0.24,-0.06)--(325:1.0);
   \foreach \a in {135,215,35,325}{\node[circle,fill=gray!55,inner sep=1.0pt] at (\a:1.0) {};}
   \node[lbl,anchor=south] at (0,1.15) {ice rule};
   \node[lbl] at (0,-1.5) {two in, two out};
 \end{scope}
\end{tikzpicture}}
\caption{Layer stacking, transfer words, and the local ice rule. Cubic and hexagonal ice
are built from the same puckered layer but use different registry sequences. Cubic stacking
gives $K^n$, alternating hexagonal stacking gives $(KK^{\mathsf T})^m$, and a mixed
sequence gives a general word in $K$ and $K^{\mathsf T}$. The lower row states the comparison proved here. In the right-hand panel, an arrow points toward the oxygen to which
the proton is covalently bound, so every oxygen has two arrows in and two out. Horizontal
offsets are exaggerated and are not crystallographic coordinates}
\label{fig:stacking}
\end{figure}

For a crystal with $N$ oxygen sites, let $Z_N$ be the number of proton configurations satisfying
the ice rule. The configuration constant $w=\lim_{N\to\infty}Z_N^{1/N}$, when it exists, is
its thermodynamic growth factor, and the configurational entropy per mole is $R\ln w$. Series,
Monte Carlo, and tensor-network calculations place the constants of both ideal endpoints near $1.5074$
\cite{Nagle1966,Kolafa2014,XuLinZhang2026}, with modern estimates agreeing to five or six
decimal places. These calculations provide compelling numerical evidence, but they necessarily
use finite systems and extrapolation. A rigorous result certifies what holds in the
infinite-crystal thermodynamic limit, independently of simulation-cell size and assumed
finite-size convergence.

This distinction leads to two mathematical problems. The first is structural: which ice-I
stacking permits the greatest number of proton configurations? Previous rigorous comparisons
treated the two ideal endpoints. The ordering $w(\Ic)\le w(\Ih)$ is due to Onsager; Li et al.\
recount it and recover it by representing cubic ice with a layer operator $M$ and a hexagonal
bilayer with $MM^{\mathsf T}$ \cite{OnsagerDupuis1960,LiebWu1972,LiCenWangYang2024}. A mixed
polytype generates a general word in $M$ and $M^{\mathsf T}$, so the endpoint comparison does
not classify arbitrary stacking sequences. Our new structural step is the identification of
every uniform-registry ice-I polytype with such a transfer word. Applying the standard
Schatten--H\"older inequality to those words proves that alternating hexagonal stacking
maximizes the ice-rule count at every matched finite cross-section and, consequently, maximizes
the thermodynamic configuration constant among all periodic uniform-registry ice-I polytypes.
Thus the novelty is the ice-polytype transfer-word identification and the certified entropy
bounds below, rather than a new matrix inequality or the introduction of $M$ and
$MM^{\mathsf T}$.
This establishes that $\Ih$ is a maximizer, not necessarily the unique maximizer. In particular,
whether $w(\Ic)=w(\Ih)$ or $w(\Ic)<w(\Ih)$ remains open; recent tensor-network calculations find
the two values numerically indistinguishable, which is numerical evidence for equality but not
a proof \cite{XuLinZhang2026}.

The second problem is quantitative: how tightly can $w$ be bounded without numerical
extrapolation? The general rigorous interval was
$\tfrac32\le w\le(9/2)^{1/3}\approx1.650964$
\cite{OnsagerDupuis1960,LiebWu1972,LasVergnas1990}, while Li et al.\ obtained the
lattice-specific lower bound $w(\Ih)\ge1.501060$ \cite{LiCenWangYang2024}.
For general simple biconnected $4$-regular graphs, Bartzos and Samaris recently proved that
the number of Eulerian orientations is at most $2\,3^{N/2}$ \cite{BartzosSamaris2026}.
The resulting asymptotic ceiling $w\le\sqrt3$ applies beyond ice networks but is less
restrictive than the ice bound above. We construct exact
finite blocks that improve the lower bound and use certified transfer-matrix inequalities to
improve the upper bounds. The resulting interval is
$1.503360\le w\le1.540196$ for every periodic uniform-registry ice-I polytype, together with
$w(\Ic)\le1.527699$ and the corresponding lower- and upper-rate bracket for fixed aperiodic
sequences along the stated exhaustions. Figure~\ref{fig:bounds-compare} places these certified intervals against the previously
known rigorous bounds. They remain wider than the numerical
uncertainty and do not resolve equality between cubic and hexagonal ice; their significance is
that they are certified in the thermodynamic limit. Combined with the stacking comparison,
they show that, within the equal-weight ice-rule model, no periodic cubic or mixed polytype has
greater conditional proton entropy than $\Ih$. Stacking-sequence multiplicity and energetic,
vibrational, interfacial, defect, and kinetic effects remain separate contributions to the
observed structure \cite{Lupi2017,Hudait2016,EngelMonserratNeeds2015}. The same argument applies
to layered constraint models whose two registry operators are nonnegative transposes on a
common state space.

\begin{figure}[!tbp]
\centering
\resizebox{\textwidth}{!}{%
\begin{tikzpicture}[x=0.62cm, font=\small,
  iv/.style={line width=5pt, line cap=butt},
  tick/.style={line width=0.6pt}]
 \draw[tick] (0,-0.55) -- (15.6,-0.55);
 \foreach \v/\lab in {0/1.50, 2.5/1.525, 5/1.55, 7.5/1.575, 10/1.60, 12.5/1.625, 15/1.65}{
   \draw[tick] (\v,-0.55) -- (\v,-0.72); \node[anchor=north] at (\v,-0.76) {\lab};}
 \node[anchor=north] at (7.8,-1.32) {configurational-entropy constant $w$};
  \draw[dashed, red!70!black, line width=0.8pt] (0.746,-0.55) -- (0.746,3.62);
 \node[red!70!black, anchor=south west, align=left, font=\scriptsize] at (0.92,3.55)
   {recent numerical estimate $\approx1.50746$};
 \draw[iv, gray!45]        (0,0)      -- (15.096,0);
 \draw[iv, blue!62!black]  (0.336,1)  -- (2.770,1);
 \draw[line width=0.7pt] (2.770,0.72)--(2.770,1.28);
 \draw[iv, orange!75!black](0.336,2)  -- (4.020,2);
 \draw[line width=0.7pt] (4.020,1.72)--(4.020,2.28);
 \draw[iv, green!45!black] (0.336,3)  -- (4.020,3);
 \draw[line width=0.7pt] (4.020,2.72)--(4.020,3.28);
 \foreach \y/\lab in {0/{previous rigorous}, 1/{cubic $\Ic$}, 2/{hexagonal $\Ih$},
                      3/{periodic polytypes and stated aperiodic rates}}
   \node[anchor=east] at (-0.25,\y) {\lab};
 \node[anchor=west, gray!45!black] at (15.25,0) {\scriptsize $1.650964$};
 \node[anchor=west, blue!62!black] at (2.93,1)  {\scriptsize $1.527699$};
 \node[anchor=west, orange!75!black] at (4.18,2){\scriptsize $1.540196$};
 \node[anchor=west, green!45!black] at (4.18,3) {\scriptsize $1.540196$};
 \draw[<-,line width=0.5pt] (0.336,-0.30)--(1.6,-0.30);
 \node[anchor=west] at (1.7,-0.30) {\scriptsize common lower bound $1.503360$};
\end{tikzpicture}}
\caption{Certified intervals for the configurational-entropy constant. The previous rigorous
interval extends from Pauling's lower bound $3/2$ \cite{Pauling1935} to the girth-six upper
bound of Las Vergnas \cite{LasVergnas1990}. The present lower bound, $1.503360$, holds for
every periodic polytype and for the stated aperiodic rates. Cubic ice has the tighter upper
bound, whereas hexagonal ice and the all-stacking comparison share the second. Relative to the
previous interval, the width is reduced by a factor of $6.20$ for cubic ice and $4.098$ over
the stacking class. If the better prior hexagonal lower bound $1.501060$ of Li et al.\
\cite{LiCenWangYang2024} replaces the Pauling endpoint, the hexagonal improvement factor is
still $4.070$. Numerical estimates are shown only for orientation and are not rigorous bounds.
Dashed and solid styles distinguish the estimate from the bounds in grayscale}
\label{fig:bounds-compare}
\end{figure}

\section{Ice I stackings as transfer-matrix words}\label{sec:setup}

Let $G$ be a $4$-regular graph. An ice state is an orientation of its edges with in-degree $2$
at every vertex; for a finite graph this count is equivalently the number of Eulerian
orientations. We write $Z_Q(\sigma)$ for the ice-state count of a finite stack with transverse
section $Q$ and stacking word $\sigma$, and $Z(G)$ for that of a finite graph $G$. In the
physical lattices, vertices represent oxygen sites and edges represent hydrogen bonds, and the
Bernal--Fowler ice rule is the resulting two-in, two-out constraint. Cubic ice corresponds to
the diamond lattice, whereas hexagonal ice corresponds to the lonsdaleite lattice. Both networks
are $4$-regular, have girth $6$, and contain a $6$-cycle through every edge.

For each periodic stacking we fix a cofinal family $\{G_n\}$ of finite periodic quotients, or tori, and
write $N_n$ for the number of vertices. The rates are
\[
\underline{w}=\liminf_{n\to\infty}Z(G_n)^{1/N_n},
\qquad
\overline{w}=\limsup_{n\to\infty}Z(G_n)^{1/N_n}.
\]
These families converge in the Benjamini--Schramm sense to their corresponding infinite networks.
Each $G_n$ is $4$-regular and hence Eulerian, so the convergence result for Eulerian orientations on
Benjamini--Schramm convergent graph sequences gives $\underline{w}=\overline{w}$
\cite{BencsBorbenyiCsikvari2024}; we write $w$ for the common value. The derivations below
establish lower bounds for $\underline{w}$ and upper bounds for $\overline{w}$ independently of
this convergence result.

Throughout, a stacking sequence assigns one uniform registry to each complete puckered layer.
Laterally varying stacking domains, partial dislocations, and general three-dimensional fault
networks are outside the transfer-word model.

\section{Hexagonal stacking is extremal}\label{sec:stack}

Fix a transverse section $Q=\mathbb Z^2/\Lambda$ and a cyclic stacking word
$\sigma\in\{\pm1\}^{2m}$, where $\sigma_\ell=+1$ and $-1$ record the two chiralities
with which one puckered layer can sit on the next. The section is chosen so that the
three intra-layer neighbors of each site remain distinct. Here $K$ and $K^{\mathsf T}$
label the two handed registry steps between consecutive layers, not the absolute
$A$, $B$, and $C$ positions; constant chirality therefore gives $ABCABC\ldots$ and
alternating chirality gives $ABAB\ldots$.

Each cell contains a $U$ and a $V$ site, so a layer has $s_Q=2|Q|$ sites. Mark an edge
as selected when its arrow points from $U$ to $V$; the ice rule then requires selected
degree two at every vertex. A boundary state records the selected interlayer bonds,
one bit per cell, and the state-space dimension is $d_Q=2^{|Q|}$.
The entry $K_{a,b}$ counts the intra-layer selections compatible with lower state $a$
and upper state $b$ for one registry. Exchanging the two sublattices reverses the
intra-layer displacements and exchanges these boundary states, so the opposite
registry has the transpose matrix. This is the relation used by Li et al., written
$\tilde M=M^{\mathsf T}$ in their notation \cite{LiCenWangYang2024}.
With registry-adapted cyclic closure, which identifies the same cell labels at the
two ends of the stack,
\begin{equation}\label{eq:word}
 Z_Q(\sigma)=\operatorname{tr}\big(A_{\sigma_1}\cdots A_{\sigma_{2m}}\big),
 \qquad A_+=K,\quad A_-=K^{\mathsf T},
\end{equation}
so cubic stacking is the constant word, giving $\operatorname{tr}(K^{2m})$, and hexagonal
stacking is the alternating word, giving $\operatorname{tr}[(KK^{\mathsf T})^m]$.

For a concrete finite-section example, take $Q=2\times3$ in the original layer coordinates and order its six one-selected-bond boundary
states by the occupied cell
$(0,0),(0,1),(0,2),(1,0),(1,1),(1,2)$. The corresponding conserved-flux block of the actual
layer operator is
\begin{equation}\label{eq:example-transfer}
K_{|a|=|b|=1}=
\begin{pmatrix}
13&11&12&13&11&14\\
12&13&11&14&13&11\\
11&12&13&11&14&13\\
13&11&14&13&11&12\\
14&13&11&12&13&11\\
11&14&13&11&12&13
\end{pmatrix}.
\end{equation}
The opposite registry has the transpose block. In particular,
$(K_{|a|=1})_{12}=11\ne12=(K_{|a|=1})_{21}$, so the transpose relation is visible rather than
formal. The full $64\times64$ operator is the direct sum of flux blocks of dimensions
$1,6,15,20,15,6,1$; exact contraction gives
$\operatorname{tr}(K^2)=87\,546$ and
$\operatorname{tr}(KK^{\mathsf T})=91\,710$. This example is a finite check of the word
construction, not a thermodynamic estimate.

The generalized H\"older inequality for Schatten norms, applied with all $2m$ exponents equal to
$2m$ so that they sum reciprocally to one, gives \cite{Bhatia1997}
\begin{equation}\label{eq:holder}
\begin{aligned}
  Z_Q(\sigma)
  &=\left|\operatorname{tr}\big(A_{\sigma_1}\cdots A_{\sigma_{2m}}\big)\right|\\
  &\le \prod_{j}\big\|A_{\sigma_j}\big\|_{2m}
   =\big\|K\big\|_{2m}^{2m}\\
  &=\operatorname{tr}\big[(KK^{\mathsf T})^{m}\big]
   =Z_Q(\Ih),
\end{aligned}
\end{equation}
because $\|K^{\mathsf T}\|_{2m}=\|K\|_{2m}$ makes every factor share one Schatten norm, and the
right-hand side is exactly the alternating word. Hexagonal stacking therefore maximizes the
ice-rule count for every common transverse section and every even number of layers.

The alternating word is also the analytically accessible one. Since $KK^{\mathsf T}$ is
positive semidefinite,
$\lambda_{\max}^m\le\operatorname{tr}[(KK^{\mathsf T})^m]\le d_Q\lambda_{\max}^m$,
where $\lambda_{\max}=\rho(KK^{\mathsf T})$, its per-site trace rate converges to
$\rho(KK^{\mathsf T})^{1/(2s_Q)}$. For a general matrix, the corresponding trace root
can fall below the spectral radius through cancellation among eigenvalues.
Positive semidefiniteness is doing the work here; both registry operators are entrywise
nonnegative.

A nonzero net registry shift means that the cyclic closure includes a lateral displacement
in laboratory coordinates. Changing coordinates preserves the count only when the boundary
identification is transformed along with the layer matrices. A different physical seam,
specified by a bond-slot permutation with state-space matrix $C$, instead gives
$\operatorname{tr}(A_{\sigma_1}\cdots A_{\sigma_{2m}}C)$, which need not equal the plain
trace. Replacing the last factor by $A_{\sigma_{2m}}C$ leaves its Schatten norm unchanged,
so the same H\"older upper bound still applies.

Three consequences follow, with different scopes. \textbf{Periodic polytypes.} For a word
of period $p$, apply Eq.~\eqref{eq:holder} to $2rp$ layers, then take cofinal periodic
quotients with $r$ and both transverse dimensions growing. This gives
$w(\sigma)\le w(\Ih)$ for every periodic uniform-registry ice-I polytype.
\textbf{Fixed aperiodic sequences.} An operator-norm estimate bounds every finite stack of
a fixed word under free, registry-adapted cyclic, or permutation-seam closure, with a boundary
factor that vanishes in the per-site root independently of the section. Along the
prism-tileable exhaustions specified in Appendix~\ref{s:aperiodic}, this places any fixed
sequence inside the bracket below. The fixed $2\times3$ section of
Eq.~\eqref{eq:example-transfer} has rate $1.540988\ldots$, above that ceiling. This rules
out a uniform ceiling over all finite sections; it does not establish that tileability is
necessary for every other growing exhaustion.
\textbf{Uniqueness.} The comparison establishes that hexagonal stacking is a maximizer,
not that it is the only one, and it does not prove $w(\Ic)<w(\Ih)$.

\section{Certified entropy bounds}\label{sec:bounds}

For a periodic polytype, $w(\sigma)$ denotes its thermodynamic configuration constant. For a
fixed nonperiodic sequence, $\underline w(\sigma)$ and $\overline w(\sigma)$ denote the lower
and upper configuration growth rates along the stated exhaustions; for periodic $\sigma$ the two
coincide. The two symbols are kept distinct because neither the existence of a single
$w(\sigma)$ nor its independence of the exhaustion is proved here for a nonperiodic sequence.

\subsection{Common lower bound}

For a graph with $N$ vertices, Nagle's positive even-subgraph expansion gives the count
$Z(G)=(3/2)^N\sum_F3^{-n_2(F)}$
\cite{Nagle1966,BorbenyiCsikvari2020}, where $F$ ranges over spanning even subgraphs
and $n_2(F)$ counts vertices of degree two. Restricting the sum to
subgraphs supported inside disjoint blocks gives a lower bound that factorizes over the blocks.
Supermultiplicativity and Fekete's lemma give an asymptotic block rate, and Perron--Frobenius
theory identifies that rate with the Perron root of the accessible block-transfer component.

\begin{figure}[!htbp]
\centering
\resizebox{0.65\textwidth}{!}{%
\begin{tikzpicture}[scale=0.92, font=\small,
  site/.style={circle,fill=black,inner sep=1.0pt},
  ghost/.style={circle,fill=gray!50,inner sep=0.9pt},
  ar/.style={->,line width=0.7pt},
  blk/.style={fill=blue!10,draw=blue!55!black,line width=0.8pt},
  bnd/.style={line width=0.45pt,gray!65}]

 \begin{scope}
  \node[anchor=base] at (1.05,2.95) {(a) admissible state};
  \node[site,inner sep=1.6pt] at (1.05,1.15) {};
  \draw[ar] (0.30,1.90)--(0.90,1.30);
  \draw[ar] (0.30,0.40)--(0.90,1.00);
  \draw[ar] (1.20,1.30)--(1.80,1.90);
  \draw[ar] (1.20,1.00)--(1.80,0.40);
  \foreach \p in {(0.25,1.95),(0.25,0.35),(1.85,1.95),(1.85,0.35)}{\node[ghost] at \p {};}
  \node[anchor=north,font=\small,align=center] at (1.05,0.15)
    {two in, two out\\at every vertex};
 \end{scope}

 \begin{scope}[xshift=5.6cm]
  \node[anchor=base] at (1.55,2.95) {(b) one block $B$};
  \foreach \x in {0,...,4}{\draw[bnd] (\x*0.62+0.3,0.3)--(\x*0.62+0.3,1.54);}
  \foreach \y in {0,...,2}{\draw[bnd] (0.3,\y*0.62+0.3)--(2.78,\y*0.62+0.3);}
  \draw[blk] (0.14,0.14) rectangle (1.70,1.70);
  \foreach \x in {0,...,4}{\foreach \y in {0,...,2}{\node[site] at (\x*0.62+0.3,\y*0.62+0.3){};}}
  \node[blue!55!black,anchor=north,font=\small,align=center] at (1.55,0.15)
    {keep only subgraphs\\supported inside $B$};
 \end{scope}

 \begin{scope}[xshift=0cm,yshift=-4.8cm]
  \node[anchor=base] at (1.86,2.95) {(c) disjoint translates};
  \foreach \i in {0,1,2}{\foreach \j in {0,1}{
     \draw[blk] (\i*1.24+0.3,\j*1.24+0.3) rectangle (\i*1.24+1.42,\j*1.24+1.42);
     \foreach \x in {0,1,2}{\foreach \y in {0,1,2}{
       \node[site] at (\i*1.24+0.46+\x*0.4,\j*1.24+0.46+\y*0.4){};}}}}
  \node[blue!55!black,anchor=north,font=\small,align=center] at (1.86,0.15)
    {every vertex in exactly one block\\no block shares a vertex};
 \end{scope}

 \begin{scope}[xshift=5.6cm,yshift=-4.8cm]
  \node[anchor=base] at (2.0,2.95) {(d) resulting bound};
  \node[anchor=north,align=center,font=\small] at (2.0,2.15)
    {every term positive\\[1pt]$\downarrow$\\[1pt]
     discard all terms\\straddling two blocks\\[1pt]$\downarrow$\\[1pt]
     product of block sums\\[2pt]
     $Z(G)\ge(3/2)^N\prod_j Z_{B_j}$};
 \end{scope}
\end{tikzpicture}}
\caption{Schematic lower-bound construction. (a) Ice states obey the two-in, two-out
rule. (b) Retain all edges internal to a chosen block. (c) Partition the vertices into
blocks of the same shape; their internal bond patterns depend on the local registry word.
(d) Discarding the positive Nagle terms that use inter-block edges gives a product of
block sums $Z_{B_j}$. The certificate uses three-layer blocks with $H=4$ and $24$ sites
per transfer column. The equivalent integer weighting is given in Appendix~\ref{s:lower}}
\label{fig:lower}
\end{figure}
\FloatBarrier

Figure~\ref{fig:lower} follows this construction through. The step that makes it work is that
every term of the expansion is positive, so terms may be discarded without tracking what they
contribute. Panel (b) keeps only those even subgraphs lying entirely inside a chosen block,
here three layers with $H=4$, giving $24$ sites per transfer column; panel (c) partitions
the crystal into disjoint blocks of this shape, so
every vertex belongs to exactly one and no term is counted twice. The discarded terms are
exactly those whose subgraph straddles a block boundary. Because a bound built by discarding
positive terms can only be too small, the result in panel (d) is valid for every block, and the
retained terms factorize across blocks, so the per-site root converts their product into
a rate. For mixed stackings the internal bond pattern depends on the local three-layer word;
taking the minimum over its eight possible types gives a common floor. Enlarging the block to a union of smaller ones that retains every previously kept
edge cannot lower the restricted-sum bound. This gives a monotone improvement for nested
block restrictions without changing the positivity argument. The bound is therefore limited by
how large a block can be enumerated exactly, not by the method.
Exact rational comparison-vector inequalities then give
\[
\begin{aligned}
  w(\sigma)&\ge 1.503360395535
  &&\text{for periodic }\sigma,\\
  \underline w(\sigma)&\ge 1.503360395535
  &&\text{for fixed aperiodic }\sigma
\end{aligned}
\]
along the stated exhaustions. The construction, discarded terms, and exact certificates are described in
Appendices~\ref{s:nagle}--\ref{s:lower}.

\subsection{Lattice-specific upper bounds}

We tile the transverse torus by open prisms running parallel to the stacking axis and sever only
intra-layer bonds. Each severed-bond variable appears in exactly two prism factors, so Finner's
degree-two hypergraph H\"older inequality replaces every prism by two replicas constrained to
agree on their common boundary \cite{Finner1992}. Following the comparison-vector method of
Calkin and Wilf and of Chan, a positive rational vector then bounds the Perron root of the
resulting two-replica operator without diagonalization \cite{CalkinWilf1998,Chan2015}.

\begin{figure}[!htbp]
\centering
\resizebox{0.65\textwidth}{!}{%
\begin{tikzpicture}[scale=0.83, font=\small,
  pr/.style={fill=gray!8,draw=black,line width=0.7pt},
  rep/.style={fill=orange!12,draw=orange!65!black,line width=0.7pt},
  cut/.style={line width=1.0pt,dashed,red!70!black},
  ax/.style={->,line width=0.8pt},
  site/.style={circle,fill=black,inner sep=0.8pt}]

 \begin{scope}
  \node[anchor=base] at (1.2,2.75) {(a) open prism};
  \draw[ax] (-0.3,0.1)--(-0.3,2.4);
  \node[anchor=east,font=\small,align=right] at (-0.35,1.25) {stacking\\axis};
  \draw[pr] (0.35,0.2) rectangle (2.05,2.4);
  \foreach \y in {0.55,1.15,1.75}{
    \foreach \x in {0.7,1.2,1.7}{\node[site] at (\x,\y){};}
    \draw[gray!60,line width=0.4pt] (0.45,\y)--(1.95,\y);}
  \node[anchor=north,font=\small,align=center] at (1.2,0.1)
    {runs through every layer\\$s=2L_uL_v$ sites per layer};
 \end{scope}

 \begin{scope}[xshift=5.4cm]
  \node[anchor=base] at (1.75,2.75) {(b) severed bonds};
  \foreach \i in {0,1,2}{\draw[pr] (\i*1.15+0.2,0.4) rectangle (\i*1.15+1.0,2.4);}
  \foreach \i in {0,1}{
    \foreach \y in {0.8,1.4,2.0}{\draw[cut] (\i*1.15+1.0,\y)--(\i*1.15+1.35,\y);}
    \node[red!70!black,anchor=base,font=\small]
      at (\i*1.15+1.18,2.44) {$\eta_{\the\numexpr\i+1\relax}$};}
  \node[anchor=north,font=\small,align=center] at (1.75,0.3)
    {each severed bond lies in\\exactly two prisms};
 \end{scope}

 \begin{scope}[xshift=0cm,yshift=-5.2cm]
  \node[anchor=base] at (1.5,2.75) {(c) cut and double};
  \draw[pr] (0.15,1.40) rectangle (1.15,2.40);
  \node[font=\small] at (0.65,1.90) {$T_n(\eta)$};
  \node[font=\large] at (1.5,1.90) {$\to$};
  \draw[rep] (1.85,1.40) rectangle (2.55,2.40);
  \draw[rep] (2.65,1.40) rectangle (3.35,2.40);
  \draw[cut] (2.55,1.90)--(2.65,1.90);
  \node[anchor=north,font=\small,align=center,text width=3.9cm] at (1.75,1.30)
    {two replicas of one prism,\\glued where $\eta$ agrees\\[2pt]
     $\sum_\eta\prod_jT_n^{(j)}\!\le\!\big(\sum_\eta T_n^2\big)^{M/2}$\\[1pt]
     product-space Cauchy--Schwarz};
 \end{scope}

 \begin{scope}[xshift=5.4cm,yshift=-5.2cm]
  \node[anchor=base] at (1.4,2.75) {(d) certificate};
  \node[anchor=north,align=center,font=\small,text width=3.6cm] at (1.4,2.6)
    {$\sum_\eta T_n(\eta)^2=\operatorname{tr}R_P^{\,n}$\\[2pt]$\downarrow$\\[2pt]
     $R_Px\le U^{2ps}x,\ x>0$\\[2pt]$\downarrow$\\[2pt]
     $w\le U$\\[4pt]
     \small cubic $2ps=36=2\times18$\\
     \small hexagonal $2ps=48=2\times24$};
 \end{scope}
\end{tikzpicture}}
\caption{Upper-bound construction. (a) Open prisms run through every layer.
(b) Each severed-bond variable belongs to two distinct prisms. (c) Finner's inequality
replaces the product of boundary-conditioned prism counts by a product of $L^2$ norms.
The sum of squared boundary-conditioned counts is the count of two replicas agreeing on
their boundary: $\sum_\eta T_n(\eta)^2=\operatorname{tr}(R_P^n)$.
(d) A positive rational vector satisfying $R_Px\le U^{2ps}x$ gives $w\le U$.
The normalization is $2ps=36$ for the cubic prism and $48$ for the hexagonal prism.
The cubic symmetry quotient is justified in Appendix~\ref{lem:g8}}
\label{fig:upper}
\end{figure}
\FloatBarrier

Figure~\ref{fig:upper} follows the four steps. The load-bearing hypothesis is the one drawn in
panel (b): each severed bond must be shared by exactly two prisms. That is what makes
the inequality a degree-two one, and it fails as soon as a bond can leave a prism and re-enter
the same prism, which is why at least two prisms are needed in each transverse direction. The
fixed $2\times3$ example does not satisfy this tiling hypothesis. Panel (c) then replaces each prism by two
replicas that agree on the shared variables; each prism count is conditioned on the severed-bond assignment $\eta$. Because every
ice-rule constraint lies wholly inside one prism, matching those assignments reconstructs
the original configuration exactly. Let $R_P$ transfer two replicas over one period,
with $s$ sites per layer per prism and $p$ layers per period. If $T_n(\eta)$ is the conditioned count for $n$
longitudinal periods, then $\sum_\eta T_n(\eta)^2=\operatorname{tr}(R_P^n)$.
Panel (d) bounds the Perron root of this two-replica operator without diagonalizing it: exhibiting one
positive rational vector with $R_Px\le U^{2ps}x$ suffices, by Collatz--Wielandt, to place the
Perron root below $U^{2ps}$. With $M$ prisms, the original graph has $N=Mpsn$ sites, and Finner's inequality gives
$Z(G)\le[\operatorname{tr}(R_P^n)]^{M/2}$. Taking the per-site root therefore gives
the exponent $1/(2ps)$, where $s$ counts sites per layer per prism and $p$ counts layers
per period. Thus $2ps=36$ for the cubic prism and $48$ for the hexagonal one. This is what makes the ceiling exact arithmetic on stored rationals
rather than a numerical eigenvalue estimate.

The two lattices require different prisms. Cubic ice uses a $3\times3$ prism with an exact quotient by the
eightfold group $G_8\cong C_2^3$ generated by replica exchange, global arrow reversal, and
transposition of the spatial bit array; hexagonal ice uses the full $3\times2$ lonsdaleite
operator with no quotient. Both rectangles use strip coordinates $(x,y)$ related to
the original layer coordinates by $(u,v)=(-x-y,-y)$; their finite domains must be
transformed with the coordinates. These constructions give
\begin{equation}\label{eq:interval}
\begin{aligned}
  1.503360395535
  &\le \underline w(\sigma)
  \le \overline w(\sigma)
  \le 1.540195787172,\\
  w(\Ic)&\le1.52769873835,
\end{aligned}
\end{equation}
with exact algebraic forms in Appendices~\ref{s:lower}--\ref{s:hex}.
Figure~\ref{fig:bounds-compare} places these endpoints against the previously known rigorous
interval. Li et al.\ reported the
narrower finite-size-extrapolated numerical approximate upper bound $1.531063(10)$
\cite{LiCenWangYang2024}; the present ceiling is less precise numerically but is a
rigorous thermodynamic bound. The cubic interval is narrower by a factor of $6.20$. For hexagonal ice, the width of the previous
Pauling--Las Vergnas interval is $4.098$ times the width of the present certified interval;
even if the stronger Li et al.\ lower bound $1.501060$ replaces Pauling's $3/2$, the
corresponding improvement factor is $4.070$. Reporting both baselines separates the gain from
the choice of earlier lower endpoint.

These certificates improve the earlier rigorous interval while leaving the numerical estimates intact.
Pauling's $3/2$ \cite{Pauling1935} is an approximate count that happens to be a rigorous lower
bound; Onsager and Dupuis established it as such and proved $w(\Ih)\ge w(\Ic)$ by relating the
two layer transfers \cite{OnsagerDupuis1960}, an ordering the present comparison recovers and
extends to every uniform-registry word. The girth-six upper bound of Las Vergnas \cite{LasVergnas1990} applies to any $4$-regular
graph of that girth and so cannot separate polytypes. Both endpoints of the previous rigorous
interval are therefore lattice-independent, which is precisely what the present certificates
give up in order to gain lattice specificity. Among numerical work, Kolafa's Monte Carlo
estimate \cite{Kolafa2014} and the finite-size extrapolation of Li et al.\
\cite{LiCenWangYang2024} are both far tighter than any certified interval, including this one;
they are not bounds, and the tensor-network calculation of Xu, Lin and Zhang
\cite{XuLinZhang2026} reports cubic and hexagonal values that are numerically
indistinguishable. Nothing here contradicts those estimates. What changes is that the
enclosing interval is now certified in the thermodynamic limit and is lattice-specific, so a
future proof of $w(\Ic)=w(\Ih)$ or of strict inequality must live inside it. Although the two certified ceilings satisfy
$1.527699<1.540196$, their order does not determine the order of the underlying entropy
constants, and the independent inequality $w(\Ic)\le w(\Ih)$ is consistent with both.

\section{Physical implications and open questions}\label{sec:discussion}

The finite-volume comparison proves that hexagonal stacking maximizes the conditional proton
entropy among periodic uniform-registry ice-I polytypes. This entropy is distinct from the
Shannon entropy of the stacking sequence itself. Within the equal-weight ice-rule model the
proton contribution $-TS_{\mathrm p}$ is therefore minimized by $\Ih$, possibly nonuniquely.
Energetic, vibrational, interfacial, defect, and stacking-sequence contributions to the
free energy remain outside this comparison, as do kinetic effects on the observed stacking \cite{Malkin2015,HartHansenKuhs2018}.

The certified interval of Eq.~\eqref{eq:interval} limits the proton configurational-entropy
spread between periodic
polytypes, and between the lower and upper rates of fixed aperiodic sequences along the stated
exhaustions, to
$R\ln(1.540195787172/1.503360395535)\approx0.2013$~J\,mol$^{-1}$\,K$^{-1}$. The bounds do not
determine whether $w(\Ic)=w(\Ih)$, whether the maximizer is unique, or whether cubic stacking
minimizes the thermodynamic entropy. Xu, Lin, and Zhang obtain a nearly normal transfer operator
and numerically indistinguishable cubic and hexagonal entropies \cite{XuLinZhang2026}. Their
calculation supplies numerical evidence compatible with equality, not a proof of
$w(\Ic)=w(\Ih)$. Wang and Zhang characterize equality
conditions for finite trace words in $K$ and $K^{\mathsf T}$ \cite{WangZhang1995}. Larger
retained blocks and replica prisms provide systematic routes to tighter certified bounds.

The proof uses only the trace-word representation of Eq.~\eqref{eq:word}, with nonnegative
transpose-related registry operators. It therefore provides an extremal criterion for other layered constraint models once
that operator structure has been established.

\section*{Statements and Declarations}
\subsection*{Competing interests}
The authors have no competing interests to declare that are relevant to the content of this article.

\subsection*{Data and code availability}
Verification code, exact certificates, production operator builders, reference outputs,
and reproduction instructions are available in the versioned Zenodo archive (v1.0.1),
\url{https://doi.org/10.5281/zenodo.21753932}~\cite{ChenRanIceData2026}.

\begin{appendices}
\renewcommand{\theHequation}{appendix.\arabic{equation}}
\renewcommand{\thefigure}{\arabic{figure}}
\renewcommand{\thetable}{\arabic{table}}
\setcounter{figure}{4}
\section{Ice I geometry and the transfer-word representation}\label{s:geom}

The oxygen networks of cubic and hexagonal ice are the diamond and lonsdaleite nets. Both are
four-coordinated and both are built from the same puckered hexagonal layer; they differ only
in the stacking registry, $ABCABC\ldots$ against $ABAB\ldots$. The two nets are locally
indistinguishable at short range, so they are separated here by their coordination sequences.

\subsection{Coordination sequences}\label{lem:cs}
The coordinate graphs have the coordination sequences
\[
  4,12,24,42,64,92 \quad\text{(diamond)},\qquad
  4,12,25,44,67,96 \quad\text{(lonsdaleite)}.
\]
The sequences first differ at graph distance three, confirming that the coordinate
constructions represent distinct nets.

\subsection{Layer representation and coordinate frames}\label{lem:g1}

Use a triangular in-plane basis with squared norm $u^2-uv+v^2$, and measure height in
bilayer-repeat units, so the three-dimensional squared norm is
\[
 \|(u,v,z)\|^2=u^2-uv+v^2+\frac23z^2.
\]
Let $\tau=(2/3,1/3)$ and $t_{\ell+1}-t_\ell=\sigma_\ell\tau$, with $t_0=0$.
The registry of the crystallographic bilayer is
$r_\ell=\sum_{j<\ell}\sigma_j\pmod3$, corresponding to $A,B,C$.
Its two planar sublattices have positions $(c+t_\ell,\ell)$ and
$(c+t_\ell,\ell+3/4)$ for $c=(u,v)\in\mathbb Z^2$.

A transfer layer groups the upper sublattice of crystallographic bilayer $\ell$ with the
lower sublattice of bilayer $\ell+1$:
\[
 U_\ell(c)=(c+t_\ell,\ell+3/4),\qquad
 V_\ell(c)=(c+t_{\ell+1},\ell+1).
\]
Thus its two sublattices have different lateral origins. For $\sigma_\ell=+1$, the three
closest $U_\ell$ cells to $V_\ell(c)$ are $c+d$ with
$d\in S_+=\{(0,0),(1,0),(1,1)\}$: each has in-plane squared separation $1/3$ and vertical
squared contribution $1/24$. For $\sigma_\ell=-1$, these displacements are negated.
Write $S_-=-S_+$.
The fourth neighbor is $U_{\ell+1}(c)$, vertically above $V_\ell(c)$, with squared
separation $(2/3)(3/4)^2=3/8$, the same as the three sloping bonds.
The nearest-neighbor rule is therefore exactly
\[
 V_\ell(c)\sim U_\ell(c+d)\quad(d\in S_{\sigma_\ell}),
 \qquad V_\ell(c)\sim U_{\ell+1}(c).
\]
This identifies the crystallographic network with the layer graph edge by edge.
It uses a regrouping of sublattices and their separate origins; translating both
endpoints of a whole layer by a common $t_\ell$ would not change its internal
displacements.

The finite blocks and replica prisms use a second, integer coordinate frame. Define
\begin{equation}\label{eq:strip-frame}
 (u,v)=(-x-y,-y),\qquad (x,y)=(v-u,-v).
\end{equation}
This unimodular map preserves the cell lattice and the vertical interlayer bonds.
The intra-layer displacements become
\[
 \widehat S_+=\{(0,0),(-1,0),(0,-1)\},\qquad
 \widehat S_-=-\widehat S_+.
\]
A rectangle $0\le x<L_u$, $0\le y<L_v$ in this strip frame is a parallelogram in
the original $(u,v)$ frame. The finite domains must be transformed with the coordinates;
we use strip-frame rectangles for the certificates throughout Appendices~\ref{s:perron}--\ref{s:hex}.

\section{Comparison across stacking words}\label{s:stack}

\subsection{Stacking-word representation and closure}\label{lem:word}

The row-to-column convention of Section~\ref{sec:stack} gives, for prescribed boundary
states $a_0,\ldots,a_n$, the product
$\prod_{\ell=1}^n(A_{\sigma_\ell})_{a_{\ell-1},a_\ell}$.
Summing the interior states gives
$M_\sigma=A_{\sigma_1}\cdots A_{\sigma_n}$.
Identifying $a_n=a_0$ therefore gives the plain trace in Eq.~\eqref{eq:word}.
In the embedding of Appendix~\ref{lem:g1}, this registry-adapted closure identifies
corresponding cells after the full displacement $(t_n-t_0,n)$.
It is a pure vertical closure when this lateral shift vanishes in the chosen transverse
quotient; otherwise the shift is included in the seam.

More generally, let $C$ be the state-space permutation induced by a bijection of the
terminal and initial interlayer bond slots, with
$C_{b,a}=1$ when the terminal state $b$ is identified with the initial state $a$.
Then
\[
 Z_{Q,C}(\sigma)=\sum_{a,b}(M_\sigma)_{a,b}C_{b,a}
 =\operatorname{tr}(M_\sigma C).
\]
A relabelling $U_\ell$ of the state basis at each interface transforms the layer
matrices and the seam together:
\[
 \widetilde A_\ell=U_{\ell-1}^{-1}A_{\sigma_\ell}U_\ell,\qquad
 \widetilde C=U_n^{-1}CU_0.
\]
The forward product now telescopes in the stated order,
\[
 \widetilde A_1\cdots\widetilde A_n=U_0^{-1}M_\sigma U_n,
 \qquad
 \operatorname{tr}(\widetilde A_1\cdots\widetilde A_n\widetilde C)
 =\operatorname{tr}(M_\sigma C).
\]
These are two coordinate descriptions of the same glued graph.
Changing the physical seam while holding the operators fixed is a different operation:
in general $\operatorname{tr}(M_\sigma C)\ne\operatorname{tr}(M_\sigma)$.

For even $n=2m$, any such permutation seam still obeys the same upper comparison.
Apply the Schatten bound of Section~\ref{sec:stack} with its last factor replaced by
$A_{\sigma_n}C$. Orthogonality of $C$ preserves the Schatten norm and gives
\[
 Z_{Q,C}(\sigma)\le\operatorname{tr}[(KK^{\mathsf T})^m].
\]
The right-hand side is the alternating count with the registry-adapted seam.
Equality of these upper bounds does not imply equality of counts with different seams.

\subsection{Why the transpose relation matters}\label{s:transpose}

The transpose hypothesis is necessary for the general comparison. For example,
\[
 A=\begin{pmatrix}0&0\\0&1\end{pmatrix},\qquad
 B=\begin{pmatrix}0&1\\1&0\end{pmatrix}
\]
are nonnegative but satisfy
$\operatorname{tr}(ABAB)=0<1=\operatorname{tr}(A^2B^2)$.
For transpose-related factors, the alternating word instead attains the common
Schatten bound. Equality conditions for finite trace words are discussed by
Wang and Zhang \cite{WangZhang1995}.

\subsection{Fixed disordered sequences}\label{s:aperiodic}

The certified interval applies directly to a fixed stacking sequence without periodic
approximants. For an infinite word
$\sigma$ and an exhaustion $(Q_k,n_k)$ with $n_k\to\infty$ and $Q_k$ running through the
prism-tileable cofinal family of strip-frame tori (side lengths divisible by
$3$ and $2$, respectively, with at least two prisms in each direction),
\[
  \underline w(\sigma):=\liminf_{k\to\infty}Z_{Q_k,n_k}(\sigma)^{1/(n_ks_{Q_k})},
  \qquad
  \overline w(\sigma):=\limsup_{k\to\infty}Z_{Q_k,n_k}(\sigma)^{1/(n_ks_{Q_k})},
\]
where $Z$ is the count under free, registry-adapted cyclic, or permutation-seam closure. No single $w(\sigma)$ is
asserted for a nonperiodic sequence: neither existence of the limit nor independence of the
exhaustion is proved here. Fix a
transverse-periodic cross-section $Q$ with $s_Q$ sites per layer, and write
$P_n=A_{\sigma_1}\cdots A_{\sigma_n}$ for the first $n$ letters. Free boundaries sum the
dangling interlayer bonds at the first and last layers over both orientations. With the seam matrix $C$ defined in Appendix~\ref{lem:word}, the three
closure counts are
\[
  Z^{\mathrm{free}}_n=\mathbf 1^{\mathsf T}P_n\mathbf 1,\qquad
  Z^{\mathrm{cyc}}_n=\operatorname{tr}P_n,\qquad
  Z^{C}_n=\operatorname{tr}(P_nC),
\]
and each is at most $d_Q\|P_n\|_{\mathrm{op}}$: the first because
$\|\mathbf 1\|_2^2=d_Q$, the second because $|M_{ii}|=|e_i^{\mathsf T}Me_i|\le\|M\|_{\mathrm{op}}$ for every square
matrix, and the third because $|\operatorname{tr}(P_nC)|\le\|P_n\|_1\le
d_Q\|P_n\|_{\mathrm{op}}$ with $C$ orthogonal. Submultiplicativity with
$\|K_Q\|=\|K_Q^{\mathsf T}\|$ gives
$\|P_n\|_{\mathrm{op}}\le\|K_Q\|_{\mathrm{op}}^n=\rho(K_QK_Q^{\mathsf T})^{n/2}$. One
interlayer bond passes per cell, so $d_Q=2^{s_Q/2}$ and the boundary factor enters the
per-site root as $d_Q^{1/(ns_Q)}=2^{1/(2n)}$, independently of $Q$. Hence for every fixed
word, every closure, and every $Q$,
\[
  \limsup_{n\to\infty}\big(Z_n\big)^{1/(ns_Q)}
  \ \le\ \rho\big(K_QK_Q^{\mathsf T}\big)^{1/(2s_Q)} ,
\]
the right side being the fixed-$Q$ alternating rate. Since $K_QK_Q^{\mathsf T}$ is positive
semidefinite,
$\lambda_{\max}^m\le\operatorname{tr}[(K_QK_Q^{\mathsf T})^m]\le d_Q\lambda_{\max}^m$, so the
alternating trace-rate limit exists and equals
$\rho(K_QK_Q^{\mathsf T})^{1/(2s_Q)}$, whereas for a general matrix
$\operatorname{tr}(M^m)^{1/m}$ can miss the spectral radius through cancellation.

For every cross-section tileable by at least two replica prisms in each transverse
direction, the certified ceiling bounds this rate. The prism inequality confines the
alternating torus counts at fixed $Q$, so
$\rho(K_QK_Q^{\mathsf T})^{1/(2s_Q)}\le\tfrac{1540195787172}{10^{12}}$. The tileability
restriction specifies the proved family: on the fixed $2\times3$ cross-section
in Section~\ref{sec:stack}, the rate is $1.540988\ldots$, above the ceiling.
This excludes a uniform statement over all finite sections, not every other growing
van Hove exhaustion. Because the boundary
correction $2^{1/(2n)}$ is uniform in $Q$, any exhaustion $(Q_k,n_k)$ with $n_k\to\infty$
through tileable cross-sections places the upper rate of the fixed sequence at or below
$1.540195787172$.

The same floor applies to all three boundary conventions. Cyclic and permutation-seam closures
produce finite $4$-regular graphs, so the positive block restriction applies directly, with incomplete boundary blocks treated as described in Appendix~\ref{s:lower}. For free
boundaries,
\[
  Z_n^{\mathrm{free}}=\mathbf 1^{\mathsf T}P_n\mathbf 1
  \ \ge\ \operatorname{tr}(P_nC)=Z_n^{C},
\]
because the trace selects a subset of the nonnegative matrix entries summed by the free count.
The same lower rate therefore follows for a fixed aperiodic sequence. Together with the upper
estimate above, this proves the bracket along the specified prism-tileable van Hove
exhaustions. Extending it to arbitrary transverse shapes would require a separate proof of
exhaustion independence.

\section{Nagle expansion and block restriction}\label{s:nagle}

\subsection{Nagle identity}\label{lem:nagle}
For a finite $4$-regular graph $G$ on $N$ vertices, with $Z(G)$ its number of ice states,
\[
  Z(G)=\Big(\tfrac32\Big)^{N}\sum_{F}3^{-n_2(F)} ,
\]
the sum running over spanning subgraphs $F$ in which every vertex has degree $0$, $2$ or $4$,
and $n_2(F)$ counting vertices of degree exactly $2$.
The identity follows from expanding the four-spin ice indicator at each vertex. Every term is
positive, so restricting the sum to any subfamily gives a lower bound; the empty subgraph
alone returns Pauling's $(3/2)^N$.

\subsection{Block restriction}\label{lem:block}
Partition the vertices of a finite $4$-regular graph into blocks $B_j$, and choose a
retained edge set $E_j\subseteq E(G[B_j])$ inside each block. Define
\[
 Z_{B_j}=\sum_{\substack{F\subseteq E_j\\\deg_F(v)\ {\rm even}}}3^{-n_2(F)},
 \qquad
 P_{B_j}=\sum_{\substack{F\subseteq E_j\\\deg_F(v)\ {\rm even}}}
 3^{\,\#\{v\in B_j:\deg_F(v)\in\{0,4\}\}}.
\]
Because $Z_{B_j}=3^{-|B_j|}P_{B_j}$, positivity and factorization give
\[
 Z(G)\ge\left(\frac32\right)^N\prod_jZ_{B_j}
       =2^{-N}\prod_jP_{B_j}.
\]
The factors may differ with the local registry word. For blocks of common size $b$,
$Z(G)^{1/N}\ge\frac12(\min_jP_{B_j})^{1/b}$; repeated identical blocks give
the familiar factor $\frac12P_B^{1/|B|}$.
All vertices must be covered. Uncovered sites require their own contribution and cannot
be omitted from the normalization.

When every internal edge is retained, the discarded terms are exactly those using an
inter-block edge. If further internal edges are removed, terms using those edges are
discarded as well. The construction below retains every internal lattice edge.

\section[Transfer operators and the Perron rate]{Transfer operators, the Fekete limit, and the Perron rate}\label{s:perron}

\subsection{Retained block and transfer entries}\label{lem:transfer}
Use the strip frame of Eq.~\eqref{eq:strip-frame}, writing its coordinates as $(r,q)$,
with transfer along $q$. For $\tau=(\tau_0,\ldots,\tau_{n-1})$, define
\[
 B(\tau,H,L)=\{U_\ell(r,q),V_\ell(r,q):
 0\le\ell<n,\ 0\le r<H,\ 0\le q<L\}.
\]
Retain exactly the lattice edges with both endpoints in this set. They are the
intra-layer bonds
\[
 V_\ell(r,q)\sim U_\ell(r,q),\
 U_\ell(r-\tau_\ell,q),\
 U_\ell(r,q-\tau_\ell)
\]
and the interlayer bonds $V_\ell(r,q)\sim U_{\ell+1}(r,q)$ whenever both endpoints remain
inside the block. Thus $|B|=2nHL$. Translates of the vertex domain partition the lattice;
different slabs inherit their own local word $\tau$.

A column has $2nH$ vertices. Its internal edge set $E_{\rm col}$ consists of the
same-column intra-layer bonds and the interlayer bonds with $0\le\ell<n-1$.
Each column vertex has one remaining possible bond across a $q$ frontier.
There are $nH$ slots on each frontier, indexed by $\ell H+r$.
For $\tau_\ell=+1$, the left slot belongs to $V_\ell(r)$ and the right slot to
$U_\ell(r)$; for $\tau_\ell=-1$, these assignments are exchanged.

For $D\subseteq E_{\rm col}$, let $d_D(v)$ be its selected degree at $v$.
Even total degree forces the vertex's frontier bit to be $d_D(v)\bmod2$.
This determines two masks $a(D),b(D)\in\{0,1\}^{nH}$ in the slot order just specified.
The transfer entry is the explicit integer sum
\[
 (T_\tau)_{a,b}
 =\sum_{\substack{D\subseteq E_{\rm col}\\a(D)=a,\ b(D)=b}}
 3^{\,\#\{v:d_D(v)\in\{0,3,4\}\}}.
\]
The exponent is the Nagle integer weight: a total degree of $0$ or $4$ contributes $3$,
and a total degree of $2$ contributes $1$. Gluing equal frontier masks and setting the two
outer masks to zero gives
\[
 P_L(\tau):=P_{B(\tau,H,L)}=(T_\tau^L)_{00}.
\]
The support digraph has an arrow $a\to b$ when $(T_\tau)_{a,b}>0$.
Write $C_\tau$ for the strongly connected component containing the empty state $0$,
and $T_{\tau,C_\tau}$ for its restriction. These definitions specify the operator
independently of its implementation.

\subsection{Perron rate}\label{prop:fekete}
For fixed $\tau$ and $H$, abbreviate $P_L=P_L(\tau)$ and $T_C=T_{\tau,C_\tau}$.
Two selected subgraphs on $L_1$ and $L_2$ columns, both returning to the empty frontier,
concatenate to an admissible configuration on $L_1+L_2$ columns. Distinct pairs give
distinct configurations, so $P_{L_1+L_2}\ge P_{L_1}P_{L_2}$.
Since $P_L\ge1$, Fekete's lemma gives the existence of $\lim_LP_L^{1/L}$.
Only states in $C_\tau$ contribute to a path from $0$ back to $0$.
The restricted matrix $T_C$ is irreducible, and the empty internal subgraph gives a
self-return at $0$. Perron--Frobenius theory therefore gives
\[
 \lim_{L\to\infty}P_L^{1/L}
 =\lim_{L\to\infty}(T_C^L)_{00}^{1/L}=\rho(T_C).
\]
Whenever a Perron root of $T_\tau$ is used below, it means this restricted component.

\section{Exact lower-bound certificates}\label{s:lower}

The Perron root is bounded below by an exact rational Collatz--Wielandt certificate
\cite{CalkinWilf1998,Chan2015}: for a
nonnegative integer matrix $T_C$ and a positive rational vector $x$,
$\rho(T_C)\ge\min_i(T_Cx)_i/x_i$, and the minimum is evaluated in exact arithmetic.
For every periodic ice-I polytype, and for the lower rate of every fixed aperiodic sequence
along the exhaustions of Sec.~\ref{s:aperiodic},
\[
  w(\sigma)\ge L_*,\qquad \underline w(\sigma)\ge L_*,
\]
where
\[
  \begin{aligned}
  L_*&:=\frac12\left(\frac{126129906455359205952}{423237029}\right)^{1/24}\\
     &>\frac{1503360395535}{10^{12}}=1.503360395535 .
  \end{aligned}
\]

The normalization follows from Appendix~\ref{lem:block}:
$Z_B=3^{-|B|}P_B$ and hence
\[
 \frac32Z_B^{1/|B|}=\frac12P_B^{1/|B|}.
\]
The integer transfer entries in Appendix~\ref{lem:transfer} use the $P_B$ convention.
A column has $2nH$ sites, so its Perron bound is converted to a per-site floor by
$\frac12\rho(T_\tau)^{1/(2nH)}$.

Table~\ref{tab:lower-certificates} lists the eight records in
\texttt{lower\_bound\_n3\_H4.json} of archive v1.0.1~\cite{ChenRanIceData2026}.
The word $\tau$ identifies the corresponding record. Each record uses a positive
rational vector on the $2048$-state component containing $0$, within a full
$4096\times4096$ transfer matrix, and supplies an entrywise inequality
$T_Cx_\tau\ge\lambda_\tau x_\tau$.

\begin{table}[!tbp]
\caption{Certificates for three-layer blocks with $H=4$ ($24$ sites per column).
The final column gives rounded-down per-site floors}\label{tab:lower-certificates}
\centering\small
\begin{tabular}{lll}
\toprule
word $\tau$ & rational comparison value $\lambda_\tau$ & per-site floor\\
\midrule
$+++$ & $126129906455359205952/423237029$ & $1.503360395535$\\
$++-$ & $203001507976071282752/681144343$ & $1.503364101402$\\
$+-+$ & $92487282677992959744/310311187$ & $1.503367631877$\\
$+--$ & $203001507976071282752/681144343$ & $1.503364101402$\\
$-++$ & $940701122202761017920/3156396493$ & $1.503364101402$\\
$-+-$ & $92487282677992959744/310311187$ & $1.503367631877$\\
$--+$ & $940701122202761017920/3156396493$ & $1.503364101402$\\
$---$ & $126129906455359205952/423237029$ & $1.503360395535$\\
\bottomrule
\end{tabular}
\end{table}
\FloatBarrier

Write $r(\sigma)=w(\sigma)$ for a periodic polytype and
$r(\sigma)=\underline w(\sigma)$ for a fixed aperiodic sequence. Decompose the stacking into
slabs of $n$ layers. Each slab carries one of $2^{n}$ block types according to its chiralities, so
$r(\sigma)\ge\tfrac12\big(\min_\tau\rho(T_\tau)\big)^{1/(2nH)}$, the minimum taken over
all types. This minimum covers every grouping offset and reflected word. At $n=3$, $H=4$,
all eight types are certified separately; the minimum Perron bound, attained by the pure cubic
block, gives the common floor.

Each slab in an arbitrary concatenation belongs to the same set of eight certified types.
The eight positive vectors differ, so the vector inequalities cannot simply be multiplied;
the product is taken over the scalar restricted sums instead. Write $P_L(\tau)=(T_\tau^L)_{00}$.
For each type, $P_L(\tau)^{1/L}\to\rho(T_\tau)$, so given $\varepsilon>0$ there is
$L_\tau(\varepsilon)$ beyond which $P_L(\tau)^{1/L}\ge\rho(T_\tau)-\varepsilon$. Because there
are only eight types, $L_0(\varepsilon)=\max_\tau L_\tau(\varepsilon)$ is finite and the
estimate holds simultaneously for all of them. The scalar sums then multiply across disjoint
slabs,
\[
  \prod_jP_L(\tau_j)\ \ge\
  \Big[\min_\tau\big(\rho(T_\tau)-\varepsilon\big)\Big]^{L N_{\mathrm{slabs}}} ,
\]
and taking the per-site root and letting $\varepsilon\downarrow0$ gives the stated floor.
Changing the grouping offset only changes which three-layer words appear. For fixed $L$, partition each large transverse section into $H\times L$ rectangles
and the stack into three-layer slabs. Assign leftover vertices to singleton blocks,
each contributing $3/2$ to the restricted count. Their volume fraction tends to zero
as all section dimensions and the layer number grow. Take this exhaustion limit first,
then let $L\to\infty$; the finite set of block types makes the preceding estimate uniform.

\section{Replica inequality and the cubic upper certificate}\label{s:cubic}

The construction does not slice along a plane. It severs only intra-layer bonds and applies a
product-space Cauchy--Schwarz inequality to the severed-bond variables, and the same argument
covers both lattices.

\subsection{Prism-tiling bound}\label{lem:cut}
Use the strip coordinates $(x,y)$ of Eq.~\eqref{eq:strip-frame}. One prism contains,
in every layer, the cells
\[
 P_{L_u,L_v}=\{(x,y,\epsilon):0\le x<L_u,\ 0\le y<L_v,
 \ \epsilon\in\{U,V\}\}.
\]
Thus the finite domain is a rectangle in the strip frame and its image under
Eq.~\eqref{eq:strip-frame} is a parallelogram in the original frame.
Translate this domain by $L_u$ and $L_v$ in the respective strip directions.
On a torus whose side lengths are multiples of $L_u,L_v$, with at least two translates
in each direction, the resulting prisms have the following properties:
\begin{enumerate}
\renewcommand{\labelenumi}{(\roman{enumi})}
\item every vertex belongs to exactly one prism;
\item every vertex carries three intra-layer and one interlayer incidence;
\item every edge is internal to one prism or joins two distinct prisms, and every
severed edge is intra-layer.
\end{enumerate}
Indeed, the cell rectangles partition the transverse torus. The interlayer edge preserves
$(x,y)$, and the intra-layer offsets $\widehat S_\pm$ are the origin and the two unit
coordinate steps of one sign. Hence every edge leaving a prism enters an adjacent prism.
The requirement of at least two translates prevents an edge from wrapping back into its
original prism. Each layer has
\[
 \begin{aligned}
 3L_uL_v-L_u-L_v&\quad\text{internal intra-layer edges},\\
 2(L_u+L_v)&\quad\text{severed-edge incidences per prism}.
 \end{aligned}
\]
These counts are $21$ and $12$ for the cubic $3\times3$ prism, and $13$ and $10$ for the
hexagonal $3\times2$ prism. Each prism has $s=2L_uL_v$ sites per layer.

Let $T_n^{(j)}(\eta_j)$ be the count on prism $j$ over $n$ longitudinal periods,
conditioned on the severed-edge bits $\eta_j$ incident to that prism.
Every local ice-rule constraint belongs to exactly one prism and sees all four bond
variables, including its boundary ones. Agreement on shared bits therefore gives a
bijection between compatible prism configurations and configurations of the whole graph:
\[
 Z(G_n)=\sum_\eta\prod_{j=1}^{M}T_n^{(j)}(\eta_j).
\]
Every severed-edge variable occurs in two distinct factors. Finner's hypergraph
H\"older inequality with exponent $2$ for every factor \cite{Finner1992} gives
\[
 Z(G_n)\le\prod_{j=1}^{M}\|T_n^{(j)}\|_2
 =\left(\sum_\eta T_n(\eta)^2\right)^{M/2}.
\]
The last equality uses congruent prisms with the same periodic stacking.

For a single layer of chirality $\sigma$, let $N_\sigma(a,b;\eta)$ count internal
assignments with lower state $a$, upper state $b$, and severed-edge data $\eta$.
All bits use the selected-edge convention of Section~\ref{sec:stack}.
The two-replica layer matrix has explicit entries
\[
 (R_\sigma)_{(a_1,a_2),(b_1,b_2)}
 =\sum_\eta N_\sigma(a_1,b_1;\eta)N_\sigma(a_2,b_2;\eta).
\]
For a period $(\sigma_1,\ldots,\sigma_p)$, put
$R_P=R_{\sigma_1}\cdots R_{\sigma_p}$. Matching successive interlayer states in each
replica and closing the stack gives
\[
 \sum_\eta T_n(\eta)^2=\operatorname{tr}(R_P^n).
\]
This is the sum of squared boundary-conditioned counts: the replicas share the severed
bits but have independent internal configurations.

The original graph has $N=Mpsn$ sites. Consequently,
\[
 Z(G_n)^{1/(Mpsn)}
 \le\operatorname{tr}(R_P^n)^{1/(2psn)}
 \le\big[\dim(R_P)\rho(R_P)^n\big]^{1/(2psn)}.
\]
The finite dimension prefactor disappears as $n\to\infty$, giving
$\overline w\le\rho(R_P)^{1/(2ps)}$ without asserting trace-root convergence.
For cubic ice $p=1$, $s=18$, and $R_P=R_+$, so $2ps=36$.
For hexagonal ice $p=2$, $s=12$, and $R_P=R_+R_-$, so $2ps=48$.

For cubic ice the open $3\times3$ prism has a two-replica operator with $262\,144$ states,
reduced to $33\,536$ by an eightfold symmetry quotient. A positive rational vector on the
quotient satisfies
\[
  \bar Rx\le\left(\frac{30553974767}{20000000000}\right)^{36}x,
  \qquad\text{whence}\qquad
  w(\Ic)\ \le\ \frac{30553974767}{20000000000}=1.52769873835 .
\]
The lift proved in Sec.~\ref{lem:g8} transfers this inequality to the full state space without
relaxation, establishing the displayed rational ceiling exactly. A potentially sharper
algebraic form $\big(\max_i(\bar Rx)_i/x_i\big)^{1/36}$ exists, but that exact maximum ratio was
not recorded when the certificate was produced, so it is not reported here.

\subsection[Exact G8 reduction]{Exact $G_8$ reduction}\label{lem:g8}
For the cubic prism, encode a one-replica boundary state by
$a=\sum_{x,y=0}^{2}a_{x,y}2^{3y+x}$, with bit $1$ denoting a selected interlayer bond.
In the embedding of Appendix~\ref{lem:g1}, this arrow points downward from $U$ to $V$.
A pair $(a_1,a_2)$ has integer index $a_1+512a_2$; the same convention applies to rows
and columns of the $2^{18}$-state operator.
Let $G_8$ act on two-replica boundary states as the group generated by replica exchange,
global arrow reversal, and transposition of the $3\times3$ spatial bit array. The three
involutions commute, so $G_8\cong C_2^3$. The full open $3\times3$ cubic replica operator
commutes with $G_8$; consequently the orbit-constant subspace is invariant, and a positive
vector $x$ on the quotient satisfying $\bar Rx\le\mu x$ lifts to a positive vector $\tilde x$
on the full space satisfying $R\tilde x\le\mu\tilde x$, with no relaxation.
Commutation with each generator makes the orbit-constant subspace, meaning vectors constant on each $G_8$-orbit,
invariant under $R$. Lift $x$ by assigning its orbit value to every
state of that orbit. Because $R$ commutes with the group, the row sum
$\sum_{s'}R_{ss'}\tilde{x}_{s'}$ is constant as $s$ ranges over an orbit, and by construction the quotient entry
$\bar R_{[s][s']}$ is exactly that sum over the target orbit. Hence
$(R\tilde x)_s=(\bar Rx)_{[s]}\le\mu x_{[s]}=\mu\tilde x_s$ for every $s$, which is the
claimed entrywise lift. Nothing is discarded, so the quotient certificate is equivalent to a
certificate on the full space.

The commutation premise is proved generator by generator rather than assumed. Replica
exchange: every entry is
\[
  R_{(a_1,a_2),(b_1,b_2)}
  =\sum_\eta N(a_1,b_1;\eta)N(a_2,b_2;\eta),
\]
followed, for $s=(a_1,a_2)$ and $t=(b_1,b_2)$, by the target-orbit-sum quotient
\[
  \bar R_{[s],[t]}=\sum_{t'\in[t]}R_{s,t'}.
\]
Here $N$ is the single-replica layer count given interlayer states $a,b$ and severed-bond data
$\eta$, and the two factors of each summand enter symmetrically. Global arrow reversal:
reversing every arrow fixes the position of every bond, complements every orientation bit,
and exchanges incoming with outgoing at every site. The ice rule, two of four incoming, is
self-dual under that exchange. Reversal is therefore a bijection of interior assignments at
complemented boundary data,
\[
  N(a,b;\eta)=N(\bar a,\bar b;\bar\eta) .
\]
The arguments do not swap: $a$ and $b$ label bonds below and above the layer, positions that
reversal leaves fixed. They are not incoming and outgoing states of a directed transfer. It follows that
\[
\begin{aligned}
R[C(s),C(t)]
&=\sum_\eta N(\bar a_1,\bar b_1;\eta)N(\bar a_2,\bar b_2;\eta)\\
&=\sum_\eta N(a_1,b_1;\bar\eta)N(a_2,b_2;\bar\eta)
=R[s,t].
\end{aligned}
\]
Since the $\eta$-sum runs over all values, $R$ commutes with the complement acting identically
on rows and columns. Spatial transposition $(x,y)\to(y,x)$ fixes the in-layer offset $(0,0)$ and exchanges
$(-1,0)$ with $(0,-1)$. It is therefore an automorphism of the open-prism layer graph,
preserving sublattices, the interior and severed-bond split, and the interlayer slots under
the same $3y+x$ bit permutation used in the group action. Hence
$N(a,b;\eta)=N(\pi a,\pi b;\pi_e\eta)$ and $R$ commutes. The three involutions commute
pairwise, which covers all eight elements. The generator actions are listed in
Table~\ref{tab:g8-actions}.

\begin{table}[htbp]
\caption{Actions of the three generators of $G_8$}\label{tab:g8-actions}
\centering\small
\begin{tabular}{llll}
\toprule
generator & rows $(a_1,a_2)$ & columns $(b_1,b_2)$ & $\eta$ and $R_{st}$\\
\midrule
exchange $X$ & $(a_2,a_1)$ & $(b_2,b_1)$ & $\eta$ fixed; factors swap\\
reversal $C$ & $(\bar a_1,\bar a_2)$ & $(\bar b_1,\bar b_2)$ & $\eta\to\bar\eta$ in the sum\\
transpose $T$ & bits $3y{+}x\to3x{+}y$ & same & slots $x{=}0\!\leftrightarrow\!y{=}0$,
$x{=}2\!\leftrightarrow\!y{=}2$\\
\bottomrule
\end{tabular}
\end{table}
\FloatBarrier

Rows and columns transform by the same permutation in every case, which is why each
generator is a similarity by a permutation matrix and never a transpose.
On the $2^{18}$ two-replica states, the complete orbit histogram is $64$ size-two orbits,
$1\,440$ size-four orbits, and $32\,032$ size-eight orbits. These sum to $33\,536$ orbits,
both by direct enumeration and by Burnside's lemma, which is the quotient dimension. The
open $3\times3$ prism layer has $21$ internal intra-layer edges and $12$ severed-edge
incidences, matching the tiling hypotheses. The quotient entries are target-orbit sums, as required
by the lift.
Combining this with Sec.~\ref{lem:cut}, the quotient certificate stated above holds on the full
space, which establishes the cubic ceiling.

\section[Lonsdaleite and the hexagonal certificate]{Lonsdaleite construction and the hexagonal upper certificate}\label{s:hex}

The hexagonal replica operator is constructed directly from the lonsdaleite layer graph, rather
than from a diamond-lattice quotient. Its prisms are columns parallel to the stacking axis, and
the severed bonds lie within layers rather than in a plane transverse to the stack. The
$3\times2$ prism satisfies the hypotheses of Section~\ref{lem:cut} with $s=12$ and $p=2$, so
$w(\Ih)\le\rho(R_h)^{1/48}$.

A positive rational vector $x>0$ with $R_hx\le\mu_h x$ entrywise gives
$\rho(R_h)\le\mu_h$ by Collatz--Wielandt, where
\[
  \mu_h=\frac{4073409405598488714800477}{4039367497556880},
  \qquad R_hx\le\mu_h x ,
\]
and therefore
\[
\begin{aligned}
  w(\Ih)&\le\mu_h^{1/48}
  =\left(\frac{4073409405598488714800477}{4039367497556880}\right)^{1/48}\\
  &<\frac{1540195787172}{10^{12}}=1.540195787172 .
\end{aligned}
\]
The exact algebraic bound is $\mu_h^{1/48}\approx1.5401957871710166$; the twelve-decimal
fraction is an outward-rounded form, certified by the exact rational inequality
\[
  \mu_h<\left(\frac{1540195787172}{10^{12}}\right)^{48}.
\]
The three certificates do not all have the same shape. The common lower bound and the
hexagonal ceiling are exact algebraic roots of stored rationals, whereas the cubic ceiling was
checked directly against a rational power and is therefore already exact as reported.

\end{appendices}

\setlength{\bibsep}{0.6em}
\bibliography{jsp-refs}
\end{document}